\documentclass[12pt]{article}

\newcommand{\beq}{\begin{equation}}
\newcommand{\eeq}{\end{equation}}
\newcommand{\beqa}{\begin{eqnarray}}
\newcommand{\eeqa}{\end{eqnarray}}
\newcommand{\nn}{\nonumber}

\def\ifmath#1{\relax\ifmmode#1\else$#1$\fi}
\def\to{\rightarrow}

\def\gsim{{~\raise.15em\hbox{$>$}\kern-.85em
          \lower.35em\hbox{$\sim$}~}}
\def\lsim{{~\raise.15em\hbox{$<$}\kern-.85em
          \lower.35em\hbox{$\sim$}~}}

\def\B{{\cal B}}
\def\L{{\cal L}}
\def\sd{\!\!/}
\def\s{\!\!\!/}

\def\L{\Lambda_{\rm QCD}}

\def\parl{{/\!/}}
\def\B{{\rm B}}
\def\D{{\rm D}}
\def\R{{\rm R}}
\def\SP{{^1S_0}}
\def\SV{{^3S_1}}
\def\DV{{^3D_1}}
\def\tr{{\rm Tr}}
\begin{document}


\title{\bf\Large Heavy-to-Light Form Factors in the\\
Final Hadron Large Energy Limit:\\ Covariant Quark Model Approach}
\author{J. Charles, A. Le Yaouanc, L. Oliver, O. P\`ene and J.-C. Raynal}
\date{January 22, 1999}
\maketitle

\begin{center}
Laboratoire de Physique Th\'eorique et Hautes \'Energies~\footnote{Laboratoire
associ\'e au Centre National de la Recherche Scientifique - URA D00063\\
E-mail: {\tt Jerome.Charles@th.u-psud.fr}, {\tt
Alain.Le-Yaouanc@th.u-psud.fr}}
\\
Universit\'e de Paris-Sud, B\^atiment 210, F-91405 Orsay Cedex, France
\end{center}

\setcounter{page}{1}

\vskip 5 mm
\begin{flushright}
LPTHE-Orsay 98-78\\ {\tt hep-ph/9901378}
\end{flushright}

\begin{abstract}
We prove the full covariance of the heavy-to-light weak current matrix
elements based on the Bakamjian-Thomas construction of relativistic quark
models, in the heavy mass limit for the parent hadron and the large energy
limit for the daughter one. Moreover, this quark model representation of the
heavy-to-light form factors fulfills the general relations that were recently
argued to hold in the corresponding limit of QCD, namely that there are only
three independent form factors describing the $B\to\pi(\rho)$ matrix elements,
as well as the factorized scaling law $\sim\sqrt{M}z(E)$ of the form factors
with respect to the heavy mass $M$ and large energy $E$. These results
constitute another good property of the quark models {\it \`a la}
Bakamjian-Thomas, which were previously shown to exhibit covariance and
Isgur-Wise scaling in the heavy-to-heavy case.
\end{abstract}

\newpage

From the point of view of the extraction of the weak couplings in $B$ decay,
and particularly $|V_{ub}|$, the knowledge of the heavy-to-light inelastic
form factors is of special interest. In addition, these form factors
constitute quite complicated hadronic objects, which contain valuable
information on the dynamics of the strong interactions. Therefore it is
important to make progress in this field.

Using the Heavy Quark Effective Theory (HQET) and the Large Energy Effective
Theory (LEET) to describe the soft contribution (i.e. the Feynman mechanism)
to the heavy-to-light weak current matrix elements, we have recently
shown~\cite{LEET} that there are only three independent form factors to
describe the pseudoscalar to pseudoscalar or vector ground state transitions
in the $M\to\infty$ and $E\to\infty$ limit, where $M$ is the initial heavy
mass, and $E$ is the energy of the daughter hadron in the initial rest frame
\beq\label{E}
E=\frac{M}{2}\left(1-\frac{q^2}{M^2}+\frac{m^{\prime\,2}}{M^2}\right)\,,
\eeq
with $m^\prime$ the mass of the light daughter hadron and $q^2$ the
four-momentum transfer. Moreover, the three universal form factors obey a
factorization formula with respect to the large scales $M$ and $E$,
$\sim\sqrt{M}z(E)$. It may even happen that the dependence with respect to $E$
(and thus to $q^2$) could be derived from first principles: indeed we have
argued in Ref.~\cite{LEET} that a $\sim 1/E^2$ scaling law should follow from
the usual expectation of a $\sim(1-u)$ suppression of the wave function of the
final state, when the Feynman variable $u$ becomes close to 1. As a support to
HQET/LEET, we have checked that all these predictions are explicitly satisfied
by the Light-Cone Sum Rule expressions for the form factors, in the
$M\to\infty$ and $E\to\infty$ limit~\cite{LEET}.

These findings should have important phenomenological consequences, because
for a strict heavy-to-light ($m^\prime\ll M$) transition the phase space is
dominated by the region where the large energy limit for the final hadron
applies. However, the overall normalization of the form factors still needs a
dynamical analysis. This is a quite difficult step. By the approximate method
of QCD sum rules on the light-cone, one obtains an expression of form factors
in terms of the final hadron distribution amplitudes~\cite{LCSR}. These
distribution amplitudes are themselves calculable from the standard QCD sum
rules, through their moments~\cite{CZphysRep}. However, in the LEET situation,
these functions  are needed near $u\sim 1$, which would require the large $n$
moments while only the lowest moments are really accessible. The result is
therefore quite uncertain.

The quark model appears as an interesting complementary dynamical method. It
makes predictions in the whole kinematical range, although uncertain too. It
may give an additional intuitive insight. The quark models for form factors
are numerous. They have been reviewed in detail in Ref.~\cite{BABAR}. One must
be aware that not all are giving predictions really derived from quark model
ideas. Many are introducing phenomenological estimates for the form factors,
inspired by Vector Meson Dominance ideas or various large mass or large
momentum transfer QCD general properties, instead of deriving them from wave
function calculations. On the other hand, when one formulates models with
three-dimensional wave functions, which have the advantage of being closely
connected with the standard quark model study of spectroscopy, it is often
difficult to satisfy general properties such as covariance or asymptotic
statements, such as the Isgur-Wise scaling and the HQET/LEET relations that we
have derived in Ref.~\cite{LEET}. A very interesting trend is then represented
by the Bakamjian-Thomas class of models using three-dimensional wave
functions~\cite{B-T}, with a rather general instantaneous interaction, which
happen to be covariant in the heavy-to-heavy limit, and which also satisfy
automatically the heavy quark symmetry~\cite{B-TIW}. Quite interestingly for
the present purpose of realizing the HQET/LEET ideas in heavy-to-light
transitions, they will be shown now to present covariance in the $M\to\infty$
and $E\to\infty$ limit, and to present the corresponding relations between the
form factors, as well as the $\sim\sqrt{M}z(E)$ scaling of the latter. At the
same time, this will give additional support to these relations~\footnote{In
fact the relations have been first suggested to us by these models, before
being settled on a QCD ground.}. This class of models relies on the old,
standard, instant form of Bakamjian-Thomas (B-T) construction of relativistic
states~\cite{B-T}, which provides a well-defined procedure to boost the wave
function from a given frame to any other, keeping fixed the number of
constituents.

Note that in the heavy-to-heavy case, the B-T models have many good features:
in addition to covariance and Isgur-Wise scaling~\cite{B-TIW}, let us mention
duality sum rules and successful quantitative description of semileptonic
$B\to D,\,D^\ast,\,D^{\ast\ast}$ transitions~\cite{D**1,BTautres}. Also, the
decay constants of the heavy mesons calculated in the B-T models compare quite
well with other approaches~\cite{BTautres}.

Let us now consider the heavy-to-light case and settle the problem by
recalling the basic features of the B-T representation of the hadronic
states~\cite{B-T,B-TIW}. A bound state of $n$ constituents, the momenta of
which are $p_i$ ($i=1\ldots n$), is described by an internal wave function
$\phi_{s_1,\ldots,s_n}(\vec{k}_2\ldots\vec{k}_n)$ where the $s_i$ are the
spins of the constituents and the internal momenta $\vec{k}_i$
($\sum\vec{k}_i=\vec 0$) are defined by
\beq
\vec{k}_i=\B^{-1}_{\Sigma p_j}\vec{p}_i\,,
\eeq
with $\B_p$ the boost $(\sqrt{p^2},\vec 0\,)\to p$. This wave function is an
eigenstate of the mass operator $M$ and of the total spin operators
$\vec{S}^2$, $S_z$. The constituents are {\it on shell}, thus
$p_i^2=k_i^2=m_i^2$. The full wave function (with the center of mass motion)
$\Psi_{s_1,\ldots,s_n}(\vec{p}_1\ldots\vec{p}_n)$ ($\sum\vec{p}_i=\vec p$, the
momentum of the bound state) is related to
$\phi_{s_1,\ldots,s_n}(\vec{k}_2\ldots\vec{k}_n)$ by a unitary transformation,
and the Hilbert space of the $\Psi$'s is an exact representation of the
Poincar\'e group~\cite{B-T,B-TIW}.

In the B-T formalism, the $\Psi\to\Psi^\prime$ current matrix element
reads~\footnote{Contrary to Ref.~\cite{B-TIW}, we choose the usual
relativistic normalization of states, hence the factor
$\sqrt{4p^{\prime\,0}p^0}$ in front of Eq.~(\ref{wigner}).}
\beqa\label{wigner}
&&\langle\,\vec{p}\,^\prime\,|\,O\,|\,\vec{p}\,\rangle=\sqrt{4p^{\prime\,0}p^0}
\int\,\left[\,\prod_{i=2}^n\frac{d^3\vec{p}_i}{(2\pi)^3}\,\right]\,
\sqrt{\frac{\Sigma p^{\prime\,0}_j\Sigma p^0_j}{M^\prime_0M_0}}\,
\left[\,\prod_{i=1}^n\sqrt{\frac{k^{\prime\,0}_ik^0_i}{p^{\prime\,0}_ip^0_i}}\,\right]\nn\\
&&\times\sum_{s^\prime_1,\ldots,s^\prime_n}\sum_{s_1,\ldots,s_n}
{\phi^\prime}\,^{\displaystyle\ast}_{s^\prime_1,\ldots,s^\prime_n}(\vec{k}_2^\prime,\ldots,\vec{k}_n^\prime)
\left[\D^\prime_1(\R^{\prime\,-1}_1)O(\vec{p}^{\,\prime}_1,\vec{p}_1)\D_1(\R_1)\right]_{s^\prime_1,s_1}\nn\\
&&\times\left[\,\prod_{i=2}^n\D_i(\R^{\prime\,-1}_i\R_i)_{s^\prime_i,s_i}\right]\phi_{s_1,\ldots,s_n}(\vec{k}_2,\ldots,\vec{k}_n)\,.
\eeqa
In the expression above, the invariant mass $M_0$ of the quark system and the
Wigner rotations $\R_i$ are functions of the $\vec{p}_i$ as follows:
\beq
M_0=\sqrt{\left(\Sigma p_i\right)^2}\,,\ \ \ \ \ \R_i=\B_{p_i}^{-1}\B_{\Sigma
p_j}\B_{k_i}\,.
\eeq
Moreover $\D_i(\R)$ stands for the matrix of the rotation $\R$ for the spin
$s_i$, and the primed quantities refer to the final state. According to the
additivity assumption of the quark model, $O(\vec{p}^{\,\prime}_1,\vec{p}_1)$
is the matrix element of the current operator $O$ between one-particle states.

From now on, we will consider heavy-to-light transitions, with a pseudoscalar
heavy meson as the parent and a light ground state pseudoscalar $P$ or vector
$V$ meson as the daughter. According to standard quantum mechanics, the
pseudoscalar ground state is
\beq\label{P}
\left\langle\left.P\right|\right.=\left\langle\left.\SP\right|\right.\,,
\eeq
while the vector one~\footnote{Note our normalization of states,
Eq.~(\ref{norm}).} is a linear combination of $\SV$ and $\DV$,
\beq\label{V}
\left\langle\left.V\right|\right.=\left\langle\left.\SV\right|\right.+\left\langle\left.\DV\right|\right.\,.
\eeq
We write the wave functions as follows:
\beq\label{fo}
\phi^{^{2S+1}L_J}_{s_1,s_2}(\vec{k}_2)=\frac{i}{\sqrt{2}}\left[\chi^{^{2S+1}L_J}(\vec{k}_2)\,\sigma_2\right]_{s_1,s_2}\phi^{^{2S+1}L_J}(\vec{k}_2)
\eeq
where
\beq\label{chi}
\chi^{\SP}(\vec{k}_2)=1\,,\ \ \ \ \
\chi^{\SV}(\vec{k}_2)=\vec{\sigma}\cdot\vec{e}\,,\ \ \ \ \
\chi^{\DV}(\vec{k}_2)=-\frac{3}{\sqrt{2}}\left[\frac{(\vec{\sigma}\cdot\vec{k}_2)
(\vec{e}\cdot\vec{k}_2)}{\vec{k}_2^{\,2}}-\frac{1}{3}\vec{\sigma}\cdot\vec{e}\right]\,,
\eeq
($\vec{e}$ is the three-dimensional rest frame polarization vector of the
vector meson). The radial wave functions $\phi^{^{2S+1}L_J}(\vec{k}_2)$ are
required to be invariant by rotation (they depend only on $|\vec{k}_2|$) and
are normalized according to
\beq\label{norm}
\int\frac{d^3\vec{k}_2}{(2\pi)^3}\,\left|\phi^{\SP}(\vec{k}_2)\right|^2=
\int\frac{d^3\vec{k}_2}{(2\pi)^3}\,\left[\,\left|\phi^{\SV}(\vec{k}_2)\right|^2+\left|\phi^{\DV}(\vec{k}_2)\right|^2\,\right]=1\,,
\eeq
while the relative normalization of $\phi^{\SV}$ and $\phi^{\DV}$ is a
dynamical quantity which could be computed by solving the bound state equation
for the vector meson.

Following Refs.~\cite{B-TIW,D**1}, we reexpress Eq.~(\ref{wigner}) in Dirac
notation, by inserting the $2\times 2$ matrices that appear above into the
$2\times 2$ upper left block of a $4\times 4$ matrix, which is then completed
with zeros. We obtain:
\beqa\label{dirac}
&&\langle\,\vec{p}\,^\prime\,|\,O\,|\,\vec{p}\,\rangle=
\int\frac{d^3\vec{p}_2}{(2\pi)^3p_2^0}\,F(\vec{p}_2,\vec{p}\,^\prime,\vec{p}\,)\times\phi^\prime\,^{\displaystyle\ast}(\vec{k}_2^\prime)\,\phi(\vec{k}_2)\nn\\
&&\times\frac{1}{16}\tr\left[O(m_1+p\s_1)(1+u\s)(m_2+p\s_2)
\B_{u^\prime}\chi^\dagger\B_{u^\prime}^{-1}(1+u\s^\prime)(m_1^\prime+p\s_1^\prime)
\right]\,,
\eeqa
with
\beq\label{jacob}
F(\vec{p}_2,\vec{p}\,^\prime,\vec{p}\,)=2\frac{\sqrt{p^{\prime\,0}u^\prime\,^0
p^0u^0}}{p_1^{\prime\,0}\,p_1^0}\left[
\frac{p_1^\prime\cdot u^\prime}{p_1^\prime\cdot u^\prime+m_1^\prime}\,
\frac{p_1\cdot u}{p_1\cdot u+m_1}\,
\frac{p_2^\prime\cdot u^\prime}{p_2^\prime\cdot u^\prime+m_2^\prime}\,
\frac{p_2\cdot u}{p_2\cdot u+m_2}\right]^{1/2}\,.
\eeq
In Eq.~(\ref{dirac}), we have defined the unit four-dimensional vectors $u$
and $u^\prime$
\beq
u=\frac{p_1+p_2}{M_0}\,,\ \ \ \ \
u^\prime=\frac{p^\prime_1+p_2}{M^\prime_0}\,,
\eeq
and we have used the same notation $O$ for the quark current operator and the
corresponding Dirac matrix. The ``boosted'' spin wave function of the final
state $\B_{u^\prime}\chi^\dagger\B_{u^\prime}^{-1}$ reads
\beqa
&&\B_{u^\prime}(\chi^\SP)^\dagger\B_{u^\prime}^{-1}=1\,,\ \ \ \ \
\B_{u^\prime}(\chi^\SV)^\dagger\B_{u^\prime}^{-1}=\gamma_5\epsilon\sd_{u^\prime}^\ast
\frac{1+u\s^\prime}{2}\,,\label{B1}\\
&&\B_{u^\prime}(\chi^\DV)^\dagger\B_{u^\prime}^{-1}=-\frac{3}{\sqrt{2}}\gamma_5
\left[\frac{(p_2\cdot u^\prime)u\s^\prime-p\s_2}{(p_2\cdot u^\prime)^2-m_2^2}\,
p_2\cdot\epsilon_{u^\prime}^\ast-\frac{1}{3}\epsilon\sd_{u^\prime}^\ast\right]
\frac{1+u\s^\prime}{2}\,,\label{B2}
\eeqa
with
\beq
\epsilon_{u^\prime}=\B_{u^\prime}(0,\vec{e}\,)\,.
\eeq

As already stressed, Eq.~(\ref{dirac}) is not covariant in general, because of
the prefactor in Eq.~(\ref{jacob}), and because the four-dimensional vectors
$p_1$ and $p_1^\prime$, being  {\it on shell} (i.e.
$p_1^{(\prime)}\,^2=m_1^{(\prime)}\,^2$), are not covariant functions of the
integration variable $p_2$ and of the external momenta $p$ and $p^\prime$. For
example one has
\beq
\vec{p}_1=\vec{p}-\vec{p}_2\ \ \ \mbox{ but }\ \ \ p_1^0=\sqrt{\vec{p}_1^{\,2}+m_1^2}\neq p^0-p_2^0\,.
\eeq
However, {\it when the active quark dominates the kinematics}, i.e. when
$u^{(\prime)}$, $p_1^{(\prime)}$ and $p^{(\prime)}$ become {\it collinear},
then the prefactor in Eq.~(\ref{jacob}) simplifies and the full
expression~(\ref{dirac}) becomes {\it covariant}. This peculiar situation is
realized in heavy-to-heavy transitions, as it is shown explicitly in
Ref.~\cite{B-TIW}. Here we shall show that this situation is also realized for
heavy-to-light matrix elements, in the limit of heavy mass for the initial
meson and large energy for the final one.

Let us now define this limit, as well as the appropriate kinematical
variables~\cite{LEET}:
\begin{itemize}
\item The four-momentum $p$, mass $M$ and four-velocity $v$ of the initial heavy meson
\beq\label{v}
p\equiv Mv
\eeq
\item The four-vector $n$ and the scalar $E$ defined by
\beq
p^\prime\equiv En\,,\ \ \ \ \ v\cdot n\equiv 1.
\eeq
Thus
\beq\label{defE}
E=v\cdot p^\prime
\eeq
is just the energy of the light meson in the rest frame of the heavy meson.
Recall the relation between $E$ and the four-momentum transfer
$q^2=(p-p^\prime)^2$:
\beq\label{q2}
q^2=M^2-2ME+m^{\prime\,2}\ \ \Longleftrightarrow\ \
E=\frac{M}{2}\left(1-\frac{q^2}{M^2}+\frac{m^{\prime\,2}}{M^2}\right)\,.
\eeq
\end{itemize}
The limit of heavy mass for the initial meson and large energy for the final
one is defined as
\beq\label{limite}
(\L,m^\prime)\ll (M,E)\,,\ \ \ \ \ \mbox{with $v$ and $n$ fixed,}
\eeq
where $\L$ in the quark model stands for the typical size of the potential.
Note that we do not assume anything for the ratios $E/M$ and $\L/m^\prime$. As
$n^2=m^{\prime\,2}/E^2\to 0$, $n$ becomes light-like in the above limit. In
the rest frame of $v$, with the $z$ direction along $\vec{p}\,^\prime$, one
has simply
\beq
v=(1,0,0,0)\,,\ \ \ \ \ n\simeq(1,0,0,1)\,.
\eeq
In a general frame one has the normalization conditions
\beq
v^2=1\,,\ \ \ \ \ v\cdot n =1\,,\ \ \ \ \ n^2\simeq 0\,.
\eeq

Now we would like to find the expansion of Eq.~(\ref{dirac}) in the
limit~(\ref{limite}). In agreement with the HQET/LEET ideas~\cite{LEET}, we
assume that the spectator quark remains {\it soft}, i.e. we consider the
limit~(\ref{limite}) with $p_2$ {\it fixed}. In addition we take $M/m_1\to 1$.
Thus we have
\beqa
p_1\simeq Mv\,,\ &&\ p_1^\prime\simeq En\,,\nn\\ M_0\simeq M\,,\ &&\
M_0^\prime\simeq\sqrt{2E\,p_2\cdot n}\,,\nn\\ u\simeq v\,,\ &&\
u^\prime\simeq\sqrt{\frac{E}{2\,p_2\cdot n}}\,n\,,\nn\\
k_2^0=\sqrt{\vec{k}_2^2+m_2^2}\simeq p_2\cdot v\,,\ &&\
k_2^{\prime\,0}=\sqrt{\vec{k}_2^{\prime\,2}+m_2^2}\simeq\sqrt{\frac{E}{2}\,p_2\cdot
n}\label{k2}\,.
\eeqa
In this limit the function $F$ in Eq.~(\ref{jacob}) becomes a Lorentz scalar
\beq
F(\vec{p}_2,\vec{p}\,^\prime,\vec{p}\,)\simeq\frac{1}{\sqrt{EM}}
\left(\frac{2E}{p_2\cdot n}\right)^{1/4}
\sqrt\frac{p_2\cdot v}{p_2\cdot v+m_2}\,.
\eeq
It remains to expand the trace in Eq.~(\ref{dirac}). As for
$\B_{u^\prime}\chi^\dagger\B_{u^\prime}^{-1}$ we have for the pseudoscalar
meson
\beq
\B_{u^\prime}(\chi^\SP)^\dagger\B_{u^\prime}^{-1}(1+u\s^\prime)
(m_1^\prime+p\s_1^\prime)\simeq En\s\,,
\eeq
while for the vector meson it is convenient to introduce, in the rest frame of
$v$ with the $z$ direction along $\vec{p}\,^\prime$, the null-vectors $N$ and
$\overline{N}$
\beq
N^\mu\equiv (1,0,0,1)\,,\ \ \ \ \ \overline{N}^\mu\equiv (1,0,0,-1)\,,
\eeq
and to specify the polarization of the final state. Indeed, denoting by
$\epsilon=\B_{p^\prime}(0,\vec{e}\,)$ the physical polarization vector, one
has
\beq\label{pol1}
\epsilon_{u^\prime}=\B_{u^\prime}(0,\vec{e}_\perp\,)=\epsilon_\perp
\ \ \ \ \ \mbox{for a transverse meson,}
\eeq
and
\beq\label{pol2}
\epsilon_{u^\prime}=\B_{u^\prime}(0,0,0,1)=u^\prime-(u^\prime\cdot N)\overline{N}
\ \ \ \ \ \mbox{for a longitudinal one.}
\eeq
Inserting Eqs.~(\ref{pol1}) and ~(\ref{pol2}) in Eqs.~(\ref{B1})
and~(\ref{B2}) and noting that $u^\prime\cdot N={\cal O}(1/\sqrt{E})$, it is
not difficult to find in the limit~(\ref{limite}), for a transverse meson
\beqa
&&\mbox{$S$-wave:}\ \ \ \ \
\B_{u^\prime}(\chi^\SV)^\dagger\B_{u^\prime}^{-1}(1+u\s^\prime)
(m_1^\prime+p\s_1^\prime)\simeq E\gamma_5\epsilon\sd_\perp^\ast n\s\,,\\
&&\mbox{$D$-wave:}\ \ \ \ \
\B_{u^\prime}(\chi^\DV)^\dagger\B_{u^\prime}^{-1}(1+u\s^\prime)
(m_1^\prime+p\s_1^\prime)\simeq
\frac{1}{\sqrt{2}}E\gamma_5\epsilon\sd_\perp^\ast n\s\,,
\eeqa
and for a longitudinal meson
\beqa
&&\mbox{$S$-wave:}\ \ \ \ \
\B_{u^\prime}(\chi^\SV)^\dagger\B_{u^\prime}^{-1}(1+u\s^\prime)
(m_1^\prime+p\s_1^\prime)\simeq E\gamma_5n\s\,,\\ &&\mbox{$D$-wave:}\ \ \ \ \
\B_{u^\prime}(\chi^\DV)^\dagger\B_{u^\prime}^{-1}(1+u\s^\prime)
(m_1^\prime+p\s_1^\prime)\simeq -\frac{2}{\sqrt{2}}E\gamma_5n\s\,.
\eeqa

As for the argument of the wave functions in Eq.~(\ref{dirac}), we have from
Eq.~({\ref{k2})
\beqa
|\vec{k}_2|&\simeq&\sqrt{(p_2\cdot v)^2-m_2^2}\,,\\
|\vec{k}_2^{\,\prime}|&\simeq&\sqrt{\frac{E}{2}\,p_2\cdot n}\,.
\eeqa
The wave functions, being invariant by rotation, are functions of the above
Lorentz scalars only.

In the end, the transition amplitude~({\ref{dirac}) becomes in the
limit~(\ref{limite})
\beqa\label{dirac2}
\langle\,\vec{p}\,^\prime\,|\,O\,|\,\vec{p}\,\rangle&=&
\sqrt{ME}\int\frac{d^3\vec{p}_2}{(2\pi)^3p_2^0}\,
\left(\frac{2E}{p_2\cdot n}\right)^{1/4}
\sqrt\frac{p_2\cdot v}{p_2\cdot v+m_2}\nn\\
&\times&\frac{1}{8}\tr\left[O(1+v\s)(m_2+p\s_2)\Gamma n\s\right]\nn\\ &\times&
\phi^\prime\,^{\displaystyle\ast}\left(\sqrt{(E/2)(p_2\cdot n)}\,\right)\,\phi\left(\sqrt{(p_2\cdot v)^2-m_2^2}\,\right)\,,
\eeqa
where the Dirac structure $\Gamma$ refers to the final state:
\beqa\label{gamma'}
&&\Gamma^{\SP}=1\,,\ \ \ \ \
\Gamma^{{\SV}_\perp}=\gamma_5\epsilon\sd_\perp^\ast\,,\ \ \ \ \
\Gamma^{{\DV}_\perp}=\frac{1}{\sqrt{2}}\gamma_5\epsilon\sd_\perp^\ast\,,\nn\\
&&\Gamma^{{\SV}_\parl}=\gamma_5\,,\ \ \ \ \
\Gamma^{{\DV}_\parl}=-\frac{2}{\sqrt{2}}\gamma_5\,.
\eeqa
As announced, Eq.~(\ref{dirac2}) is {\it covariant}, which is another
remarkable feature of the B-T formalism.

Thanks to its covariance properties, Eq.~(\ref{dirac2}) can be further
reduced. One defines three form factors $A$, $B$ and $C$ by
\beqa
A&=&\sqrt{ME}\int\frac{d^3\vec{p}_2}{(2\pi)^3p_2^0}\,
\left(\frac{2E}{p_2\cdot n}\right)^{1/4}
\sqrt\frac{p_2\cdot v}{p_2\cdot v+m_2}\,\frac{m_2}{8}\nn\\
&\times&
\phi^\prime\,^{\displaystyle\ast}\left(\sqrt{(E/2)(p_2\cdot n)}\,\right)\,\phi\left(\sqrt{(p_2\cdot v)^2-m_2^2}\,\right)\,,\label{A}\\
B\,v^\mu+C\,n^\mu&=&\sqrt{ME}\int\frac{d^3\vec{p}_2}{(2\pi)^3p_2^0}\,
\left(\frac{2E}{p_2\cdot n}\right)^{1/4}
\sqrt\frac{p_2\cdot v}{p_2\cdot v+m_2}\,\frac{p_2^\mu}{8}\nn\\
&\times&
\phi^\prime\,^{\displaystyle\ast}\left(\sqrt{(E/2)(p_2\cdot n)}\,\right)\,\phi\left(\sqrt{(p_2\cdot v)^2-m_2^2}\,\right)\,.\label{BC}
\eeqa
Then the matrix element~(\ref{dirac2}) reads
\beqa
\langle\,\vec{p}\,^\prime\,|\,O\,|\,\vec{p}\,\rangle&=&\tr\left[O(1+v\s)(A+Bv\s+Cn\s)\Gamma n\s\right]\\
&=&(A+B)\times\tr\left[O(1+v\s)\Gamma n\s\right]\,.\label{dirac3}
\eeqa
Indeed $(1+v\s)v\s=(1+v\s)$ and from Eq.~(\ref{gamma'}), $n\s$ commutes or
anticommutes with $\Gamma$, which yields
$n\s\,\Gamma\,n\s=\pm\Gamma\,n\s\,^2\simeq 0$.

To summarize, we consider for definiteness the $B\to P(V)$ transitions, and
from Eqs.~(\ref{P}), (\ref{V}), (\ref{A}), (\ref{BC}) and~(\ref{dirac3}) we
obtain
\beqa
\langle\,P\,|\,O\,|\,B\,\rangle&=&2E\,\zeta(M,E)\,
\tr\left[\frac{1}{4}O(1+v\s)n\s\right]\,,\\
\langle\,V_\parl\,|\,O\,|\,B\,\rangle&=&2E\,\zeta_\parl(M,E)\,
\tr\left[-\frac{1}{4}O(1+v\s)n\s\gamma_5\right]\,,\\
\langle\,V_\perp\,|\,O\,|\,B\,\rangle&=&2E\,\zeta_\perp(M,E)\,
\tr\left[\frac{1}{4}O(1+v\s)n\s\gamma_5\epsilon\sd_\perp^\ast\right]\,,
\eeqa
where the form factors $\zeta$, $\zeta_\parl$ and $\zeta_\perp$ are given in
terms of the overlap integrals $\zeta^{^{2S+1}L_J}$
\beqa
\zeta(M,E)&=&\zeta^\SP(M,E)\,,\label{f+}\\
\zeta_\parl(M,E)&=&\zeta^\SV(M,E)-\frac{2}{\sqrt{2}}\,\zeta^\DV(M,E)\,,\\
\zeta_\perp(M,E)&=&\zeta^\SV(M,E)+\frac{1}{\sqrt{2}}\,\zeta^\DV(M,E)\,,\label{ff1}
\eeqa
with
\beqa
\zeta^{^{2S+1}L_J}(M,E)&=&\frac{\sqrt{M}}{4}\int\frac{d^3\vec{p}_2}{(2\pi)^3p_2^0}\,
\left(\frac{2}{E\,p_2\cdot n}\right)^{1/4}
\sqrt\frac{p_2\cdot v}{p_2\cdot v+m_2}\,\left(m_2+p_2\cdot n\right)\nn\\
&\times& {\phi^{^{2S+1}L_J}}^{\displaystyle\ast}\left(\sqrt{(E/2)(p_2\cdot
n)}\,\right)\,\phi_B\left(\sqrt{(p_2\cdot v)^2-m_2^2}\,\right)\,.\label{ff2}
\eeqa
Calculating explicitly the traces for
$O=1\,,\gamma_5\,,\gamma^\mu\,,\gamma^\mu\gamma_5\,,
\sigma^{\mu\nu}\,,\sigma^{\mu\nu}\gamma_5$, we recover the HQET/LEET prediction derived in
Ref.~\cite{LEET}, namely that {\it there are only three independent form
factors describing the transitions $B\to P(V)$ in the limit of heavy mass for
the initial meson and large energy for the final one}:
\beqa
\left\langle P\left|V^\mu\right|B\right\rangle&=&
2E\,\zeta(M,E)n^\mu\,,\label{res1Deb}\\
\left\langle P\left|T^{\mu\nu}\right|B\right\rangle&=&
i2E\,\zeta(M,E)\left(n^\mu v^\nu-n^\nu v^\mu\right)\,,\\
\left\langle V\left|V^\mu\right|B\right\rangle&=&
i2E\,\zeta_\perp(M,E)\,\epsilon^{\mu\nu\rho\sigma}v_\nu
n_\rho\epsilon^\ast_\sigma\,,\\
\left\langle V\left|A^\mu\right|B\right\rangle&=&
2E\left\{\zeta_\perp(M,E)\left[\epsilon^{\ast\mu}-(\epsilon^\ast\cdot
v)n^\mu\right] +\,\zeta_\parl(M,E)\frac{m_V}{E}(\epsilon^\ast\cdot
v)n^\mu\right\}\,,\\
\left\langle V\left|T_5^{\mu\nu}\right|B\right\rangle&=&
-i2E\zeta_\perp(M,E)\left(\epsilon^{\ast\mu}n^\nu-
\epsilon^{\ast\nu}n^\mu\right)\\
&&-i2E\zeta_\parl(M,E)\frac{m_V}{E}(\epsilon^\ast\cdot v)\left(n^\mu
v^\nu-n^\nu v^\mu\right)\,,\label{res1Fin}
\eeqa
where $V^\mu=\overline{q}\gamma^\mu b$, $A^\mu=\overline{q}\gamma^\mu\gamma_5
b$, $T^{\mu\nu}=\overline{q}\sigma^{\mu\nu}b$ and
$T_5^{\mu\nu}=\overline{q}\sigma^{\mu\nu}\gamma_5 b$ are respectively the
vector, axial, tensor and pseudotensor weak currents, with $q$ the appropriate
light quark field.

Moreover, Eqs.~(\ref{ff1})-(\ref{ff2}) exhibit a {\it factorized scaling law}
$\sim\sqrt{M}\,z(E)$ with respect to the large scales $M$ and $E$, which is
also in agreement with our general results~\cite{LEET}. Note that contrary to
the heavy-to-heavy case~\cite{B-TIW}, the form factors depend on the final
state through the wave functions $\phi^{^{2S+1}L_J}$, and there is no definite
normalization of the form factors at some particular $q^2$.

We would like to make some additional comments. While the dependence with
respect to the heavy mass $M$ of the form factors requires no dynamical
analysis, the knowledge of their dependence with respect to the large energy
$E$ (and thus to $q^2$) needs the study of the behaviour of the internal wave
function $\phi(\vec{k})$ at large relative momentum $|\vec{k}|$ (see
Eq.~(\ref{ff2})), that depends {\it a priori} on the specific potential of the
constituent quark model. We have shown explicitly in Ref.~\cite{LEET} that the
Light-Cone Sum Rules, plus the assumption of the asymptotic-like behaviour for
the distribution amplitudes near $u\sim 1$, predict a definite $1/E^2$ scaling
law, and we have argued that the latter dependence might be a very general
property of QCD. The question of whether the quark model with a QCD-inspired
potential can agree with this result is left for further investigation.

Our approach {\it \`a la} Bakamjian-Thomas is similar in spirit with the ones
of Stech~\cite{stech} and Soares~\cite{soares}. These authors have considered
a general constituent quark picture and have obtained relations between the
form factors which are close to Eqs.~(\ref{res1Deb})-(\ref{res1Fin}), provided
that $\zeta_\parl=\zeta_\perp$. As already said in Ref.~\cite{LEET}, we have
found no general reason supporting the latter identity; this even seems in
contradiction with the Light-Cone Sum Rule approach, which does not imply
$\zeta_\parl=\zeta_\perp$. In the B-T formalism, Eqs.~(\ref{ff1})-(\ref{ff2}),
$\zeta_\parl=\zeta_\perp$ would hold if the $D$-wave were asymptotically
suppressed by some power of $E$ with respect to the $S$-wave in the vector
meson; such a suppression would depend on the structure of the spin-dependent
interaction in the potential, and thus is not guaranteed {\it a priori}.

It is useful to discuss the connection of the instant form of Bakamjian-Thomas
models, that we have just described, with the particular $P=\infty$ models
which are based on a similar B-T construction on the
null-plane~\cite{terentev,polyzou}. The latter models have two versions,
unequivalent because the approach is not covariant in general, the
longitudinal~\cite{simula} and the transverse~\cite{jaus} ones, the null-plane
axis being  respectively parallel or perpendicular to the three-dimensional
momentum transfer; in the transverse case, one has to perform an analytical
continuation to go into the physical region for semileptonic
decays~\cite{jaus} \footnote{Note that the dispersion formulation of
Ref.~\cite{melikhov} is initially meant  to continue the standard~\cite{jaus}
$P=\infty$ transverse formulation (recall that it is not defined in the
physical region), but it differs in fact from the latter for certain form
factors, due to the introduction of ``subtraction terms''.}. We have
shown~\cite{unpub} that these null-plane models are recovered from the instant
form through large velocity Lorentz boosts ($P=\infty$). Now, covariance is
satisfied for the instant form in the heavy-to-heavy limit~\cite{B-TIW}, and
also, as  shown above, in the limit of heavy mass for the initial hadron and
large energy for the final one, relevant for heavy-to-light meson transitions.
Then it may be expected, by interchange of the limits with the
infinite-momentum boost, that the instant and null-plane approaches will give
the same results in these limits. Namely, the corresponding limits of the
null-plane models would also be covariant and the same for the longitudinal
and transverse cases, and identical to the covariant limit of the instant form
that we have just derived. Indications favoring this intuition are, in the
heavy-to-heavy case, the equality of the slope $\rho^2$ of the Isgur-Wise
function calculated for the same mass operator (see the discussion in the
paper on $B\to D,D^\ast,D^{\ast\ast}$ in Ref.~\cite{BTautres}), and in the
heavy-to-light case, the fact that our expression~(\ref{f+}) for $f_+$
coincides at $q^2=0$ with the null-plane formula of Ref.~\cite{jaus}.

To conclude, we have shown that the quark model representation of the
heavy-to-light transition amplitudes based on the Bakamjian-Thomas
construction of states become covariant in the heavy mass limit for the parent
hadron, and the large energy limit for the daughter one. Moreover, in
agreement with the general results that we have found in Ref.~\cite{LEET},
only three independent form factors are needed to describe all the ground
state to ground state matrix elements in this limit; these universal form
factors satisfy a simple scaling law with respect to the large scales,
$\sim\sqrt{M}\,z(E)$.

\end{document}